\documentclass{amsart}

\usepackage{amssymb}
\usepackage{amsthm}
\usepackage{amsmath,amsfonts,amsthm}
\usepackage{thmtools}
\usepackage{etoolbox}
\usepackage{mathtools}
\usepackage{soul}
\usepackage{tcolorbox}
\usepackage{times}
\usepackage{algorithm}
\usepackage{algorithmicx}
\usepackage{algpseudocode}
\usepackage{tabularx}
\usepackage{caption}
\usepackage{latexsym} 
\usepackage{fleqn}
\usepackage{tikz}
\usepackage{subfiles}
\usepackage{microtype}
\usepackage{url}

\newtheorem{definition}{Definition}

\newtheorem{lemma}[definition]{Lemma}
\newtheorem{theorem}[definition]{Theorem}

\usepackage{latexsym} 
\usepackage{fleqn}
\usepackage{tikz}
\usepackage{subfiles}

\usepackage[a4paper, total={6in, 8in}]{geometry}

\makeatletter
\newcommand{\multiline}[1]{%
  \begin{tabularx}{\dimexpr\linewidth-\ALG@thistlm}[t]{@{}X@{}}
    #1
  \end{tabularx}
}
\makeatother

\newcommand{\problemDef}[3]
{%
    \begin{tcolorbox}[arc=0.1mm,boxsep=-0.6mm,left=1.9mm,right=1.9mm,bottom=1.4mm,top=1.4mm,adjusted title={\strut \sc#1},colback=white!5]

    \noindent\textbf{Instance:} #2
    \noindent\textbf{Question:} #3
    \end{tcolorbox}
}

\newcommand*{\naturals}{\mathbb{N}}
\newcommand{\Ordo}{{\mathcal{O}}}
\newcommand{\ordo}{{o}}

\newcommand{\cdc}{\textsc{CDPA}}
\newcommand{\allen}{\textsc{Allen}}

\author{Leif Eriksson}
\address[L. Eriksson]%
   {Dep. Computer and Information
     Science, \\Link\"opings  Universitet, Sweden
   }
\email{leif.eriksson@liu.se}   
\author{Victor Lagerkvist}
\address[V. Lagerkvist]%
   {Dep. Computer and Information
     Science, \\Link\"opings  Universitet, Sweden}
\email{victor.lagerkvist@liu.se}      

\title{Improved Algorithms for Allen's Interval Algebra by Dynamic Programming with Sublinear Partitioning}

\begin{document}
\maketitle

\begin{abstract}
    {\em Allen's interval algebra} is one of the most well-known calculi in qualitative temporal reasoning with numerous applications in artificial intelligence. Recently, there has been a surge of improvements in the {\em fine-grained} complexity of NP-hard reasoning tasks, improving the running time from the naive $2^{\Ordo(n^2)}$ to $\Ordo^*((1.0615n)^{n})$, with even faster algorithms for unit intervals a bounded number of overlapping intervals (the $\Ordo^*(\cdot)$ notation suppresses polynomial factors).
    Despite these improvements the best known lower bound is still only $2^{\ordo(n)}$ (under the {\em exponential-time hypothesis}) and major improvements in either direction seemingly require fundamental advances in computational complexity. 
    In this paper we propose a novel framework for solving NP-hard qualitative reasoning problems which we refer to as {\em dynamic programming with sublinear partitioning}. 
    Using this technique we obtain a major improvement of $\Ordo^*((\frac{cn}{\log{n}})^{n})$ for Allen's interval algebra. To demonstrate that the technique is applicable to more domains we apply it to a problem in qualitative spatial reasoning, the {\em cardinal direction point algebra}, and solve it in $\Ordo^*((\frac{cn}{\log{n}})^{2n/3})$ time. Hence, not only do we significantly advance the state-of-the-art for NP-hard qualitative reasoning problems, but obtain a novel algorithmic technique that is likely applicable to many problems where $2^{\Ordo(n)}$ time algorithms are unlikely.
\end{abstract}

\section{Introduction}\label{sec:intro}

{\em Allen's interval algebra} ({\allen}) is one of the most well-known examples of qualitative temporal reasoning where the task is to decide whether a set of (typically incompletely specified) intervals in $\mathbb{R}^2$ can be ordered in a consistent way.
{\allen} is an influential formalism with a wide range of applications in artificial intelligence, and for a handful of concrete examples we may e.g.\ mention planning~\cite{DBLP:conf/ijcai/AllenK83,DBLP:journals/dke/Dorn95,DBLP:conf/icra/MudrovaH15,DBLP:conf/aaai/PelavinA87}, natural language processing~\cite{DBLP:conf/ijcai/DenisM11,DBLP:conf/aaai/SongC88}, and molecular biology~\cite{DBLP:journals/jacm/GolumbicS93}. The second major problem that we consider, the {\em cardinal direction point algebra} ({\cdc}), can be seen as an offshoot of Allen's interval algebra where the task is to reason about directions of objects in a 2-dimensional space. This algebra has seen applications in e.g. geographical information science and image retrieval and we refer the reader to the survey by Dylla et al.~\cite{Dylla:2017:SQS:3058791.3038927} for additional references and applications for both these formalisms.

Both these formalisms are in general NP-hard and admit non-trivial tractable fragments.
Hence, in actual applications one is rarely contempt with reasoning only with tractable fragments, which spurs the question of how fast we can solve NP-hard qualitative reasoning tasks. First, we need to carefully establish a baseline upper bound that we can use to measure whether an algorithm for these problems is indeed an improvement or not. This is easy for many types of problems over {\em finite} universes, e.g., finite domain {\em constraint satisfaction problems} (CSPs) and the Boolean satisfiability problem, which can always be solved simply by enumerating all possible assignments from variables to the finite universe. Clearly, this cannot be done for {\allen} or {\cdc}, but these problems can be solved in $2^{\Ordo(n \cdot \log n)}$ time by enumerating certain total orderings, which for {\allen} gives the precise upper bound $\Ordo^*((2n)^{2n})$~\cite{lagerkvist2017d}\footnote{The notation $\Ordo^*(\cdot)$ suppresses polynomial factors and $n$ denotes the number of intervals.}. Hence, the question is whether this upper bound can be improved, and we stress that this is sometimes known to be possible (e.g., $k$-SAT and $k$-COLORING) but that there also exist problems where this is conjectured to be impossible (e.g., CNF-SAT, SAT with unrestricted clause length). However, this has been proven possible for {\allen} which was recently solved in $\Ordo^*((1.0615n)^{n})$ time~\cite{eriksson2021} by a novel dynamic programming algorithm. Additionally, if {\allen} is restricted to intervals of length one, then it can even be solved in  $2^{\Ordo(n \log \log n)}$ time~\cite{DBLP:conf/kr/DabrowskiJOO20}. A faster $\Ordo^*(c^n)$ time algorithm is also known for the special case where no point occurs inside more than $k$ intervals~\cite{ijcai2022p251}. However, despite these improvements, we are still far away from an unconditional single-exponential $\Ordo^*(c^n)$ time algorithm and even further away from the best-known lower bounds which only rule out subexponential algorithms running in $2^{\ordo(n)}$ time under the {\em exponential-time hypothesis}~\cite{jonsson2021}.

In this paper we make a significant push for {\allen} and {\cdc} by introducing a new variant of dynamic programming;  {\em dynamic programming with sublinear partitioning}, which is exceptionally suited for problems related to orderings. After having introduced the necessary preliminaries (in Section~\ref{sec:prelim}) we first showcase our new method for {\allen} in Section~\ref{sec:allen} where we obtain the 
bound $\Ordo^*((\frac{cn}{\log{n}})^{n})$, for some constant $c \geq 1$.  
This is a major improvement compared to the aforementioned bound of $\Ordo^*((1.0615n)^{n})$ and thus proves that {\allen} is solvable in $\Ordo^*(f(n)^{n})$ time for $f \in \ordo(n)$, i.e., a significant leap towards a $\Ordo^*(c^n)$ time algorithm. To exemplify that our method is not limited to {\allen} we continue (in  Section~\ref{sec:cdc}) by showing how it can be tailored to solve {\cdc} faster than the general case of {\allen}. Here, we obtain the bound $\Ordo^*((\frac{cn}{\log{n}})^{2n/3})$ for a constant $c$, which is {\em much} faster than the algorithm for {\allen}.
The main idea behind our dynamic programming strategy is to use a table consisting of records, representing (partially) placed start- and end-points of intervals, where the twist is that the size of a record is not fixed beforehand but may expand up to a certain size during the execution of the algorithm.
To achieve this we use a similar idea to the one presented in Eriksson \& Lagerkvist~\cite{eriksson2021} but we analyze the structure of potential solutions much deeper.
We manage to identify redundant information by using the fact that if we know the relative position of more variables then we need to keep track of fewer potential positions.
While previous approaches have required keeping track of, or storing information corresponding to, a linear number of partitions measured in the number of variables, we here manage to solve the task while only partitioning the variables into a sublinear number of partitions.
Consequently, the upper limit of information stored in a record is bounded tighter by the number of variables than any earlier approach.
In turn, this leads to a lower number of maximum records, and hence a lower overall complexity. 
This idea of managing with only a sublinear number of partitioning will be our main result throughout the paper.
However, achieving this is as expected far from trivial.

Last, it is worth observing that virtually all qualitative temporal and spatial reasoning problems can be formulated as CSPs over a template containing  a strict partial order or an acyclic relation~\cite{jonsson2021}. Hence, all members of this diverse and rich class of problems are intrinsically related to the existence of certain orders, and it is not a large stretch of the imagination to conjecture that more problems can be solved with our approach. Most pressingly: Can dynamic programming with sublinear partitioning be used to solve an NP-hard qualitative reasoning problem in $\Ordo^*(c^n)$ time? We discuss this and additional questions in Section~\ref{sec:conc}.

\section{Preliminaries}\label{sec:prelim}

Given a set of finitary relations $\Gamma$ defined on a (potentially infinite) set $D$ of values, we define the constraint satisfaction problem over $\Gamma$ (CSP$(\Gamma)$) as follows.
\problemDef{CSP$(\Gamma)$}
{
A tuple $(V, C)$, where $V$ is a set of variables and $C$ a set of constraints of the form $R(v_1, \ldots, v_t)$, where $t$ is the arity of $R \in \Gamma$ and $v_1, \ldots, v_t \in V$.
}
{
Is there a function $f \colon V \rightarrow D$ such that $(f(v_1), \ldots, f(v_t)) \in R$ for every $R(v_1, \ldots, v_t) \in C$?
}

The set $\Gamma$ is referred to as a \emph{constraint language}, while the function $f$ is a function \emph{satisfying} the instance $I$, or simply a \emph{model} of $I$.
Given an instance $I$ of CSP$(\Gamma)$, we let $||I||$ denote the number of bits required to represent $I$.

We continue by defining total orders and order relations induced by a given total order. 
A total order $(S, \leq)$ is a relation $\leq$ on set $S$ which is {\em reflexive} (for all $x \in S$, $x \leq x$), {\em antisymmetric} (for all $x, y \in S$, if $x \leq y$ and $y \leq x$, then $x = y$), {\em transitive} (if $x \leq y$ and $y \leq z$, then $x \leq z$), and {\em strongly connected} (for all $x, y \in S$, either $x \leq y$ or $y \leq x$).
If $\odot \in \{<,>,=\}$ and $T = (S, \leq_T)$ is a total order then we write $\odot_T$ for the relation induced by $T$: $x <_T y$ if $x \leq_T y$ and $y \leq_T x$ does not hold, conversely for $>_T$, and $x =_T y$ if $x \leq_T y$ and $y \leq_T x$.
The total number of different total orders for a variable set of $n$ elements is known as $n$'th \emph{Ordered Bell Number} ($OBN(n)$) and is in $\Ordo^*((\frac{n}{e\log{2}})^n)$ and $\Ordo^*((0.5307n)^n)$ (recall that the $O^*(\cdot)$ notation supresses polynomial factors).

\subsection{Allen's Interval Algebra}

In {\em Allen's interval algebra} the basic relations consists of 13 relations between intervals over $\mathbb{R}^2$ (summarized in Table~\ref{tab:IAdesc}). The computational problem is then defined as follows.

\begin{table}
    \centering
    \begin{tabular}{l|l|l}
         $\boldsymbol{Relation}$ & $\boldsymbol{Illustration}$ & $\boldsymbol{Interpretation}$ \\
         \hline
         $\boldsymbol{X<Y}$ & XXX& X precedes Y \\
         $\boldsymbol{Y>X}$ & $\,\,\,\,\,\,\,\,\,\,\,\,\,\,\,\,\,\,$YYY & \\%Y is preceded by X \\
         \hline
         $\boldsymbol{X\,=\,Y}$ & XXXXXXX & X is equal to Y \\
          & YYYYYYY &  \\
         \hline
         $\boldsymbol{X\,m\,Y}$ & XXX & X meets Y \\
         $\boldsymbol{Y\,mi\,X}$ & $\,\,\,\,\,\,\,\,\,\,\,\,\,$YYY & \\%Y is met by X \\
         \hline
         $\boldsymbol{X\,o\,Y}$ & XXXXX & X overlaps with Y \\
         $\boldsymbol{Y\,oi\,X}$ & $\,\,\,\,\,\,\,\,\,\,\,\,\,\,$YYYY & \\%Y is overlapped by X \\
         \hline
         $\boldsymbol{X\,s\,Y}$ & XXX & X starts Y \\
         $\boldsymbol{Y\,si\,X}$ & YYYYYYY & \\%Y is started by X \\
         \hline
         $\boldsymbol{X\,d\,Y}$ & $\,\,\,\,\,\,\,\,\,$XXX & X during Y \\
         $\boldsymbol{Y\,di\,X}$ & YYYYYYY & \\%Y contains X \\
         \hline
         $\boldsymbol{X\,f\,Y}$ & $\,\,\,\,\,\,\,\,\,\,\,\,\,\,\,\,\,\,$XXX & X finishes Y \\
         $\boldsymbol{Y\,fi\,X}$ & YYYYYYY & \\%Y is finished by X \\
    \end{tabular}
    \caption{The 13 basic relations between two intervals on the same line. ($i$ denotes the $i$nverse/converse of a relation.)}
    \label{tab:IAdesc}
\end{table}

\problemDef{Allen's Interval Algebra}
{A set of intervals $V$ and a set of binary constraints $C$ of the form $c(X,Y)$ where ${c \subseteq   \{<,>,=,m,mi,o,oi,s,si,d,di,f,fi\}}$ and ${X,Y \in V}$.
}
{Is there a total order $T = (S, \leq)$ and a function $f \colon V\rightarrow S^2$ such that for all $X\in V$ with $f(X)=(u,v)$ then $u<v$ and for every constraint $c(X,Y)\in C$ then $f(X) \odot f(Y)$ for some $\odot \in c$?}

Equivalently, the problem can be defined as CSP$(\mathcal{A})$ where $\mathcal{A}$ contains the 13 basic relations in Table~\ref{tab:IAdesc} and is closed under union.
Note from Table~\ref{tab:IAdesc} that any relation between two intervals can be described using point relations ($<$, $>$ and $=$) between the start- and end-point of the two intervals. For a set of intervals $V$ we write $V^-$ for the set of start-points and $V^+$ for the set of end-points.

\subsection{Cardinal Direction Point Algebra}
We follow the projection based approach of Frank~\cite{DBLP:conf/ogai/Frank91} and define the second major problem of the paper as follows.

\problemDef{Cardinal Direction Point Algebra}
{A set of pairs of points, $V$, and a set of binary constraints $C$, of the form $c(u,v)$, where 
${c \subseteq \{\{<,>,=\} \times \{\{<,>,=\}\}}$
and ${u,v \in V}$.
}
{Is there a function $f \colon V\rightarrow \{1,\dots,|V|\}^2$ such that for every constraint $c(u,v)\in C$ then $f(u)=(u_x,u_y)$, $f(v)=(v_x,v_y)$, $u_x \odot_x v_x$ and $u_y \odot_y v_y$ with $(\odot_x,\odot_y) \in c$?}

Alternatively, the problem can be seen as whether $V$ can be ordered into two separate total orderings such that one represents the $x$-dimension and the other the $y$, such that they combined satisfy all constraints.
Another alternative is viewing {\cdc} as a sub-problem of {\allen} (e.g. by letting $a^-$ represent $a_x$ and $a^+$ represent the $a_y$ coordinate of a point $a$).

\section{A New Algorithm for Allen's Interval Algebra}\label{sec:allen}

In this section we present a novel, improved algorithm for {\allen} based on dynamic programming. The twist with this algorithm is that the dimension of our table, or the size of the records stored, is not static, but may dynamically change during the execution of the algorithm. This dynamic programming is realized by using records functioning as elements or keys in a map that is dynamically built, and we refer to this programming scheme as {\em  dynamic programming with sublinear partitioning}.
Here, sublinear refers to the fact that we partition the variable set into a sublinear number of partitions.
Each such record represents a set of $S$ variables that have been used to construct this record, and then pairs of disjoint subsets of $S$, and Booleans.
The subsets of $S$ are here used to keep track of the (relevant) ordering between variables that have been used, while the Booleans serve as markers that tell us if a certain subset has ever been merged with another subset or not.
This is useful, since if they have previously been merged with some other subset, we do not want to add any new variables to this subset, since we could then get inconsistent cases where $x=y$, $y=z$ but $x<z$.
Combined, these partitions of $S$ represent an ordered partition, but we will not strictly use them as such.

We also introduce two notations for records:
\emph{left-heaviness} where we require start-points of intervals in a record to always be in either of the first two subsets, and \emph{pair-placing} where we require whole intervals, and not just a start- or an end-point, to be placed at the same time into a record. Formally, these notions are now defined as follows.

\begin{definition}\label{def:record}
For a CSP$(\mathcal{A}$) instance $(V,C)$ we say that $r=(S,(S_1,B_1),\dots,(S_m,B_m))$ is a \emph{Record} if the following holds:
\begin{enumerate}
    \item $S \subseteq V^-\cup V^+$,
    \item $B_1,\dots,B_m \in \{0,1\}$, and
    \item $S_1,\dots,S_m$ is a partitioning of $S$.
\end{enumerate}

We say that $m$ is the size of the record.
A record is \emph{left-heavy} if $S_i\cap V^-=\emptyset$ for all $3\leq i \leq m$.
A record is \emph{pair-placing} if and only if for all $x^- \in S_i$ for any $i$, there is a $j > i$ such that $x^+\in S_j$ or $i=j$ and $B_i=0$.

For any two variables $u,v$ in some Record with ${u\in S_i}$ and ${v \in S_j}$, then we also assume that $u<v$ if $i<j$, $u>v$ if $i>j$ and that $u=v$ if $i=j$ and $B_i=1$.
However, if $i=j$ and $B_i=0$ we have to assume all three relations between $u$ and $v$ holds.
Further, a Record \emph{contradicts} $C$ if and only if there is a constraint ${c(X,Y)\in C}$ such that no assumption we can make about the relations $X$ and $Y$ given said Record will satisfy $c(X,Y)$.

\iffalse
Further, a Record \emph{contradicts} $C$ and $I$ iff there is a constraint $c(X,Y)\in C$ with $x^-\in S_i$, $x^+\in S_j$, $y^-\in S_{i'}$ and $y^+\in S_{j'}$ such that one of the following does not hold:
\begin{enumerate}
    \item if $i<i'$ then ${c(X,Y) \cap \{<,m,o,di\}\neq \emptyset}$,
    \item if $i='i$ and $B_i=1$ then ${c(X,Y) \cap \{=,s,si\}\neq \emptyset}$,
    \item if $i>i'$ then ${c(x,y) \cap \{>,mi,oi,d,f\}\neq \emptyset}$,
    
    \item if $i<j'$ then ${c(X,Y) \subseteq \{>,mi\}}$,
    \item if $i=j'$ and $B_i=1$ then ${c(X,Y) = \{mi\}}$,
    \item if $i>j'$ then ${c(X,Y) = \{>\}}$,

    \item if $j<i'$ then ${c(X,Y) \cap \{<\}\neq \emptyset}$,
    \item if $j=i'$ and $B_i=1$ then ${c(X,Y) = \{m\}}$,
    \item if $j>i'$ then ${c(X,Y) \subseteq \{<,m\}}$,

    \item if $j<j'$ then ${c(X,Y) \cap \{<,m,o,s,d\}\neq \emptyset}$,
    \item if $j=j'$ and $B_i=1$ then ${c(X,Y) \cap \{=,f,fi\}\neq \emptyset}$, or
    \item if $j>j'$ then ${c(X,Y) = \{>,mi,oi,si,di\}}$.
\end{enumerate}
\fi
\end{definition}

Generalization between records, i.e. whether a record $A$ can be converted to a record $B$ by adding variables and merging subsets, will be important, and we define the following concept.

\begin{definition}\label{def:recordCompatible}
We say that for two records \[{r_1=(S_1,(S_{1,1},B_{1,1}),\dots,(S_{1,m},B_{1,m}))}\] and \[{r_2=(S_2,(S_{2,1},B_{2,1}),\dots,(S_{2,k},B_{2,k}))}\] then $r_2$ is a generalization of $r_1$ if:
\begin{enumerate}
    \item $S_1\subseteq S_2$,
    \item $k\leq m$ ($r_2$ is not larger than $r_1$),
    \item for every $S_{1,i}$ there is a $S_{2,j}$ such that $S_{1,i}\subseteq S_{2,j}$ (no set in $r_1$ is split in $r_2$),
    \item for every $j$ then $B_{2,j}=1$ if and only if either $S_{2,j}=\emptyset$ or if there is an $i$ such that $B_{1,i}=1$ and $S_{1,i} = S_{2,j}$, otherwise $B_{2,j}=0$ (only empty or unchanged sets keep their positive $B$ value), and 
    \item for all variables $x\in S_{1,i}$ and $x'\in S_{1,i'}$ with $i\leq i'$ then $x\in S_{2,j}$ and $x'\in S_{2,j'}$ such that $j\leq j'$ (ordering is kept).
\end{enumerate}
\end{definition}

For a record ${r=(S,(S_1,B_1),\dots,(S_m,B_m))}$ we can recursively generate all generalizations $r'=(S,(S'_1,B'_1),\dots,(S'_{m'},B'_{m'}))$ with $m'\leq m$ in $\Ordo^*(3^m)$ time.
We do this by branching on every $S_i$ and constructing two new records, one where $S_i$ is merged with $S_{i-1}$ (when these exists) and $(S_{i-1},B_{i-1})$ is removed and one where we also replace $(S_{i-1},B_{i-1})$ with $(\emptyset,1)$.
For the value of $B'_i$ we strictly follow the definition.
By checking if a given result has earlier already been created we can also prevent calling the recursion with the same record more than once.
Our complexity of $\Ordo^*(3^m)$ then comes from that either ${(S'_i,B'_i)=(S_i,B_i)}$, ${(S'_{i-1},B'_{i-1})=(S_{i-1},B_{i-1})\cup (S_i,B_i)}$ or ${(S'_{i},B'_{i})=(S_{i-1},B_{i-1})\cup (S_i,B_i)}$ and ${(S'_{i-1},B'_{i-1})=(\emptyset,1)}$, creating fewer than $3^m$ different possibilities.
Let $\mathrm{GenerateGeneralizations}(r,H)$ be the function doing this for a record $r$ and where $H$ is the set of all previously generated records.
Note that the records in $H$ does not need to be generalizations of $r$, since this set is just used to prevent unessesary computational complexity.

After this we want to continue by adding new intervals to existing records.
When adding new intervals to records we do so in such a manner that given an {\allen} instance $I=(V,C)$ we never contradict $C$, while, if given a left-heavy and pair-placing record, the result is also left-heavy and pair-placing.

We briefly remind the reader that for an interval $X$ the points $x^-$ and $x^+$ represents the start- and endpoint of said interval.

\begin{algorithm}
\caption{Add a new pair to a given record $R$, assuming $B_2=1$, not contradicting $C$} 
\label{alg:recordPlace}
\begin{algorithmic}[1]
\Function{Add}{$I=(V,C)$, \newline
\hspace*{4em}$R=(S,(S_1,B_1),\dots,(S_m,B_m)), Output$}
    \For{every $x^-,x^+ \in (V^-\cup V^+)\setminus{\bigcup_{i=1}^m S_i}$ \newline 
    \hspace*{4em}and every $2\leq j \leq m$ with $B_j=1$}
        \If{for all $X'\in V$ with $x'^-\in S_u$ and $x'^+ \in S_v$ all\newline
            \hspace*{4em}$c(X,X'),c(X',X)\in C$ are satisfied by \newline
            \hspace*{4em}$x^- = 2$, $x^+ = j$, $x'^-=u$ and $x'^+=v$ }
            \State $Output \gets $GenerateGeneralizations$((\newline
            \hspace*{5em} S\cup\{x^-,x^+\},
            (S_1,B_1),(S_2\cup \{x^-\},1),\dots,\newline
            \hspace*{5em} (S_j\cup \{x^+\},1),\dots,(S_m,B_m)),Output)$
        \EndIf
    \EndFor
    \State \Return $Output$
\EndFunction
\end{algorithmic}
\end{algorithm}

As we can see, Algorithm~\ref{alg:recordPlace} guarantees that for any pair $X$ that we add to a record $(S,(S_1,B_1),\dots,(S_m,B_m))$, all constraints $c(X,X')\in C$ with $X'\in S$ are satisfied.
Quite obviously the work done in Algorithm~\ref{alg:recordPlace} (if we exclude the work done by the functio $\mathrm{GenerateGeneralizations}$) is also polynomial in terms of $||I||+m$.
By only ever adding $x^-$ to $S_2$ the algorithm keeps left-heaviness if the input is left-heavy, and as we add one pair $(x^-,x^+)$ at a time, the result is also pair-placing if the input is pair-placing.
Here we also see how the Boolean variables $B_i$ are used, as we only add new start- and end-points to sets with a Boolean set to $1$.
Combined with the definition of generalization this helps ensure we avoid situations where now $x=y$ and $x=z$, but $x<z$ in an earlier record which the current record is a generalization of.

With Algorithm~\ref{alg:recordPlace} we now have everything we need for solving {\allen}.
However, we want to optimize our method and hence want to limit our records to a certain size, i.e. limit our $m$.
The choice of $m$ needs to be big enough to ensure correctness of the algorithm, i.e. big enough so that we can always ensure that every pair can be placed into some record, while the number of already placed pairs increases.
Choosing $m=4n$ would trivially give us enough room to place every single variable, be it a start- or end-point, in any possible ordering.
To do better we first make the following observation: 
given an arbitrary ordering of $V^-\cup V^+$ and placing pairs following the ordering of the start-points, if we let $k$ be the number of pairs yet to be placed, we need at most $2k$ pairs $(S_i,B_i)$ for which $B_i=1$.
This leads to a situation where we have records where for every second pair, the Boolean is set to one while for the other pairs it is set to zero, while still allowing the placement of every remaining variable.
If more than $2k+1$ pairs with $B_i=0$ exists, there must exist two pairs $(S_j,0)$ and $(S_{j+1},0)$ and hence a generalization with $(S_j\cup S_{j+1},0)$, but which is otherwise identical. 
Hence we take $m=4k+1$, which also leads us to our main algorithm as presented in Algorithm~\ref{alg:allenMain}.

\begin{algorithm}
\caption{Solving {\allen} by dynamic programming.} 
\label{alg:allenMain}
\begin{algorithmic}[1]
\Function{Main}{$I=(V,C)$}
    \State $R\gets \{(\emptyset,(S_1=\emptyset,1),\dots,(S_{4n+3}=\emptyset,1))\}$
    \For{every record \newline
            \hspace*{4em}$r=(S,(S_1,B_1),\dots,(S_m,B_m))\in R$}
        \If{$|S|=2n$}
            \State \Return $\mathit{True}$
        \ElsIf{$m\leq 4(n-|\bigcup_{i=1}^m S_i|)+3$}
            \State $R \gets $Add$(I,r,R)$
        \EndIf
    \EndFor
    \State \Return $\mathit{False}$
\EndFunction
\end{algorithmic}
\end{algorithm}
    
In Algorithm~\ref{alg:allenMain} we see how we make use of this decreasing number of pairs $(S,B)$ by ignoring any records containing more pairs that we think we will need.
This algorithm works by iteratively calling Algorithm~\ref{alg:recordPlace} with new records, until every record has been either tested, or we find a record containing all intervals (in which case we must have a satisfiable instance).

\begin{lemma}\label{lem:MainCorr}
For an arbitrary CSP$(\mathcal{A})$ instance $I$, Algorithm~\ref{alg:allenMain} returns $\mathit{True}$ if and only if $I$ is a satisfiable instance.
\end{lemma}
\begin{proof}
We begin by proving completeness.
Given a satisfiable instance $I=(V,C)$, we know there exists a function ${f : V^-\cup V^+ \rightarrow \naturals}$ such that $f$ satisfies every constraint in $C$. Let ${V^- = \{x_1^-, \ldots, x_n^-\}}$ and ${V^+ = \{x_1^+, \ldots, x_n^+\}}$ such that ${f(x_1^-)\leq f(x_2^-)\leq \dots\leq f(x_n^-)}$, i.e., each pair $x_i^-,x_i^+$ is assigned an index $i$ according to some ordering of the start-points of the intervals. 
Define the sets ${T_1,\dots,T_{n}}$ such that ${x_j^\pm \in T_i}$ if and only if there is no variable ${v \in \{x_i^-,x_i^+,\dots,x_n^-,x_n^+\}}$ such that ${f(v)<f(x_j^\pm)}$.
Furthermore, define the auxiliary function ${\#_< : V^-\cup V^+ \times 2^{V^-\cup V^+} \rightarrow \naturals}$ such that ${\#_<(x_j^\pm, U)}$ returns the number of variables $y$ such that ${f(y)< f(x_j^\pm)}$ and $y\in U$.
From this we define the functions ${f_{\#,i} : V^- \cup V^+ \rightarrow \{1,\dots,4n+1\}}$:
{\begin{align*}
    f_{\#,i}(x_j^\pm)=&2\cdot\#_<(x_j^\pm,\{x_i^-,x_i^+,\dots,x_n^-,x_n^+\})+ \\
    &\begin{cases}
        2, &\text{ if } x_j^\pm \in T_i, \\
        1, &\text{ if } x_j^\pm \not \in T_i.
    \end{cases}
\end{align*}}
Since $f_{\#,i}$ follows the same ordering as $f$, it is also a solution for $I[\{x_i^-,x_i^+,\dots,x_n^-,x_n^+\}]$, i.e. the sub-problem of $I$ where only variables $\{x_i^-,x_i^+,\dots,x_n^-,x_n^+\}$ are considered. 
Additionally, for all intervals $Y\in V$ and $X_i\in \{T_i\cup\ldots\cup T_n\}$, $f_{\#,i}$ also satisfies any constraints $c(X_i,Y)\in C$.
For each $i\in\{0,\dots,n\}$ define the record
{\begin{align*}
    r_i=(\{x_1^-,x_1^+,\dots,x_i^-,x_i^+\},&(S_{1,i},B_{1,i}),\dots,\\
    &(S_{4(n-i)+3,i},B_{4(n-i)+3,i}))
\end{align*}} such that $x_j^\pm \in S_{i',i}$ if and only if $f_{\#,i}(x_j^\pm,)=i'$ and $j\leq i$, and $B_{i',i}=1$ if and only if $S_{i',i}\cap S_i = \emptyset$ and ${B_{i',i}=0}$ otherwise.
According to our definitions $r_{i+1}$ is a generalization of $r_i$, and every record is both left-heavy and pair-placing.
As such, Algorithm~\ref{alg:allenMain} generates all records $r_i$, ${i\in\{0,\ldots,n\}}$ starting from the original record constructed at line 2, which is equivalent to $r_0$.
By doing so the algorithm must reach a record containing all variables ($r_n$), and hence returns $\mathit{True}$.

For soundness, assume that the algorithm returns $\mathit{True}$. 
Then there must be an ordering of the variables in $V$, $X_1,\dots,X_n$, and sequence of records $r_0,\ldots,r_n$ such that
\begin{align*}
    r_i=(\{x_1^-,x_1^+,\dots,x_i^-,x_i^+\},&(S_{1,i},B_{1,i}),\dots,\\
    &(S_{4(n-i)+3,i},B_{4(n-i)+3,i}))
\end{align*}
for all ${i\in\{0,\ldots,n\}}$ and such that
\begin{align*}
    (\{x_1^-,x_1^+,\dots,x_i^-,x_i^+\},&(S_{1,j}\setminus{S'},B_{1,j}),\dots,\\
    &(S_{4(n-j)+3,j}\setminus{S'},B_{4(n-j)+3,j}))
\end{align*}
is a generalization of $r_i$ for all $j \in \{i+1,\ldots,n\}$ when
$$
S' = \{x_{i+1}^-,x_{i+1}^+,\dots,x_j^-,x_j^+\}.
$$
This holds because of how $GenerateGeneralizations$ together with $Add$ constructs records.
Since $B_{i,j}$ must be $1$ for $Add$ to add new a new pair to $S_{i,j}$, since  (1) $B_{i,j}$ is set to $0$ as soon as two sets are merged into $S_{i,j}$, and (2) every constraint $c(X_i,X_{i'})\in C$ must be satisfied for all $i'<i$, for $X_i$ to be added.
Given this set, define the function $h : V^-\cup V^+ \rightarrow \{1,\ldots, 4n+3\}$ such that (assuming $i>i'$) if $x_i^\pm \in S_{j,i}$ and $x_{i'}^\pm \in S_{j',i}$ with $j\odot j '$ then $h(x_i^\pm)\odot h(x_{i'}^\pm)$.
Since every $c(X_i,X_j)\in C$ is satisfied for all $j<i$ when $X_i$ is added (since $h$ orders $X_i$ according to its position when added), and since this is true for all $i$, $h$ must satisfy $C$.
Hence, $h$ is a solution for $I$, implying that our algorithm is sound.
With both soundness and completeness proven, the algorithm must also be correct, and hence we are done.
\end{proof}

With all supporting lemmas in place, we can now present the main result of the section.

\begin{theorem}\label{theo:mainAllen}
There is a constant $c$ such that {\allen} can be solved in $\Ordo^*((\frac{cn}{\ln{n}})^n)$ time and space using dynamic programming with sublinear partitioning.
\end{theorem}
\begin{proof}
We know that Algorithm~\ref{alg:allenMain} is correct from Lemma~\ref{lem:MainCorr}.
For the run-time we note that $Add$ is polynomial (with respect to $n$ and $m$), $GenerateGeneralizations$ adds fewer than $3^m$ sets to $Output$ and does linear work in terms of the size of the return.
Our $Main$ itself also only does polynomial work for each record, and hence the question that remains is the maximal possible number of records.
With the way we limit records size, we have that for a certain $m$ there are at most ${poly(n,m,)\cdot 2^{4(n-m)} \binom{n}{m} (8(n-m))^{m}}$ different records with $m$ pairs placed.
Here, $2^{4(n-m)}$ is an overestimation for our possible assignments of Booleans, $\binom{n}{m}$ is the different ways to choose pairs, $2^m$ is from start-points being in either $S_{1,m}$ or $S_{2,m}$, and ${(4(n-m))^{m}}$ comes from the number of ways to place end-points into the available $S_{i,m}$.
In total we hence have roughly $\sum_{0=m}^n2^{4(n-m)} \binom{n}{m} (8(n-m))^{m}$ records, which is linear in terms of the maximum value of $2^{4(n-m)} \binom{n}{m} (8(n-m))^{m}$.
Since $2^{4(n-m)} \binom{n}{m} 8^m\leq32^n$ we are left with having to maximize $(n-m)^{m}$.
Since $n,m,n-m\geq 0$ and $n,m\in \naturals$, replacing ${n-m}$ with $x$ (and $m$ with $n-x$) and derivation gives us ${\frac{d}{dx}x^{n-x}=x^{n-x-1}(n-x-x\ln{x})}$.
The maximum must then occur when $n=x+x\ln{x}$, which have no exact finite expression,
but can be estimated as $x=\frac{n}{\ln{\frac{n}{\ln{n}}}}$.
Since $n<\frac{n}{\ln{\frac{n}{\ln{n}}}}\ln{\frac{n}{\ln{\frac{n}{\ln{n}}}}}$ this gives us an overestimation of $x$, and hence we get a valid upper bound on $x^{n-x}$ by also approximating $n-x=n$.
Inserting this in the original formula and assuming $n$ is large enough gives us 
\[
\big(\frac{cn}{\ln{(n/\ln{n})}}\big)^n = \big(\frac{cn}{\ln{n} - \ln\ln{n}} \big)^n \leq \big(\frac{2cn}{\ln{n}} \big)^n.
\]
Hence, we have an upper bound on the form $\Ordo^*((\frac{cn}{\ln{n}})^n)$ where $c$ is some constant, which completes the proof.
\end{proof}

Compared to the previous best known upper bound of $\Ordo^*((1.0615n)^n)$~\cite{eriksson2021} this is a clear improvement.
In fact, we even changed the category {\allen} belongs to by pushing it below $\Ordo^*((cn)^n)$ for every constant $c>0$.
This may be a major cornerstone towards a single-exponential algorithm for {\allen}.

While the algorithm presented here only decides consitency by outputing a $\mathit{True}$ or $\mathit{False}$ answer, one can using standard implementations of e.g. backpoints also use the algorithm to yield solutions to satisfiable instances. Or even modify the algorithm to instead count solutions.

\section{Cardinal Direction Calculus}\label{sec:cdc}
In this section we focus on {\cdc}.
While {\cdc} can be reduced to a subproblem of {\allen}, the problem is interesting in its own right and is typically viewed as an independent formalism.
The {\cdc} formalism also allows significantly better complexity than the general case of {\allen}
and we will show that CDC can be solved in $O^*((\frac{cn}{\log{n}})^{2n/3})$ time for a constant $c$.

We achieve this in three steps: first we modify the algorithm from Eriksson \& Lagerkvist~\cite{eriksson2021} to work in the now two-dimensional space.
Second, we show that by running the modified version four times in different combinations of directions, while not storing orderings longer than roughly $n/3$, the worst case is significantly improved.
Last, we show how the method from Section~\ref{sec:allen} can be applied, yielding the final complexity of $\Ordo^*((\frac{cn}{\log{n}})^{2n/3})$.

Briefly, the algorithm presented in Eriksson \& Lagerkvist~\cite{eriksson2021} works on intervals by dynamic programming and adding  points in a brute force way (regardless of whether they are start-points or end-points) in sets of equality, one set of a time, to the end of the working ordering.
While the complexity of this approach is generally bounded by the size of the ordering, we can further bound the size of said ordering, since if we remove whole intervals from the ordering when both start- and end-point are placed, the ordering will never contain any end-points, except in the very last position.

For our first approach to {\cdc} we follow a similar strategy but in two dimensions:
we use dynamic programming and keep track of two active orderings, one in the direction of positive $x$-relations, and one positive $y$-relations direction.
We also keep track of points already having been placed into both orderings, and as a result we also, directly or indirectly, know which variables that have not been used in either ordering.
Effectively, we have records $(S,(S_x,\leq_x),(S_y,\leq_y))$ where $(S_x\cup S_y) \subseteq S \subseteq V$ while $S_x \cap S_y = \emptyset$ and both $(S_x,\leq_x)$ and $(S_x,\leq_x)$ are total orders.
By only placing (sets) of points into the smaller of the two orderings, and only at the very end, and then ``forgetting'' points used in both orderings, we achieve the worst case for this approach.
Again, these partitions of $S$ form an ordered partition, but we will not strictly use them as such.

\begin{lemma}\label{lem:cdc1}
{\cdc} can be solved in $\Ordo^*(2^{3n/2}OBN(n/2)^{2})$ or $\Ordo^*((0.7506n)^n)$ time with dynamic programming.
\end{lemma}

\begin{proof}
(Sketch)
Consider the records we described earlier.
Correctness, similar to how it worked for {\allen}, comes from {\cdc} only containing binary constraints.
This allows us to only check constraints once a variable has been fully placed, since we then know how said variable relates to all other variables, and hence know if the constraint is satisfied or not.
Once a constraint has been satisfied, we can thereafter forget the variable, since we will not move it inside our ordering, and hence any constraints must keep being satisfied.
For the complexity we applied the restriction that we cannot add new variables to the larger of the two orderings, unless they are equal, and we remove variables from the orderings once they are in both.
This leads to a situation where we can in the worst case have two orderings of size $n/2$ and $2^n$ ways to partition the variables between those two, leading to $O^*(2^{n}OBN(n/2)^2)$ records.
Finally, for each worst-case record there are at most $O^*(2^{n/2})$ subsets of $V$ that we can add to the end of the smaller of the two orderings.
Hence we do at most $O^*(2^{n/2})$ for each worst-case record.
For non-worst-case records, we may do more work (up to $O^*(2^n)$), but as these records are significantly fewer, the asymptotic behavior is dominated by the worst-case ones.
Hence, we have the complexity of $O^*(2^{3n/2}OBN(n/2)^{2})$ or $O^*((0.7506n)^n)$.
\end{proof}

This is already enough to surpass the previous best result for {\allen}~\cite{eriksson2021} but does not beat the bound from Section~\ref{sec:allen}.
To improve this approach for {\cdc} consider the scenario of having the solution be all points on a diagonal line from positve $X$ and negative $Y$ towards negative $X$ and positive $Y$.
This represents the worst-case for the algorithm outlined previously and naturally raises the question of whether one could order in negative $x$- or $y$-direction instead (since this would yield $\Ordo^*(1)$ complexity for this case instead). 
However, there are of course other cases (we could just mirror this example along either axis) that are instead the worst case for this approach.
So, instead we try all four possible combinations of directions: positive $x$ + positive $y$, positive $x$ + negative $y$, negative $x$ + positive $y$ and negative $x$ + negative $y$.
Using this approach and not allowing the orderings to grow larger than some maximum, we are able to avoid all these worst cases.

\begin{definition}
Given a finite total ordering $(S,<)$ and an integer $k\geq0$ let $Sub_k((S,<))$ be the first $k$ elements in $(S,<)$, or $S$ if $k\geq|S|$.
\end{definition}

\begin{figure}[t]
      \begin{tikzpicture}[scale=2.3]

        \draw[-, dashed] (-0.2,-0.2) -- (3.2,-0.2) node[right] {$x$};
        \draw[-, dashed] (-0.2,-0.2) -- (-0.2,3.2) node[above] {$y$};

    \draw[step=1cm,gray,very thin] (0,0) grid (3,3);
    \node[anchor=center] at (0.5,3.5-1) {$x_{3,1}$};
    \node[anchor=center] at (0+1.5,3.5-1) {$x_{3,2}$};
    \node[anchor=center] at (0+2.5,3.5-1) {$x_{3,3}$};

    \node[anchor=center] at (0.5,3.5-2) {$x_{2,1}$};
    \node[anchor=center] at (0+1.5,3.5-2) {$x_{2,2}$};
    \node[anchor=center] at (0+2.5,3.5-2) {$x_{2,3}$};

    \node[anchor=center] at (0.5,3.5-3) {$x_{1,1}$};
    \node[anchor=center] at (0+1.5,3.5-3) {$x_{1,2}$};
    \node[anchor=center] at (0+2.5,3.5-3) {$x_{1,3}$};
    \end{tikzpicture}
  \caption{Graphic representation of the nine sets described by $V$, $Sub_{k_{++}}(T^+_x)$, ${Sub_{k_{++}}(T^+_y)}$, $Sub_{k_{--}}(T^-_x)$ and $Sub_{k_{--}}(T^-_y)$.
  E.g.
  ${x_{1,1} = |Sub_{k_{++}}(T^+_x) \cap Sub_{k_{++}}(T^+_y)|}$, 
  ${x_{1,2} = |Sub_{k_{++}}(T^+_y) \setminus{Sub_{k_{++}}(T^+_x)\cup Sub_{k_{--}}(T^-_x)}|}$, 
  ${x_{1,3} = |Sub_{k_{++}}(T^+_y) \cap Sub_{k_{--}}(T^-_x)|}$ and 
  ${x_{2,1} = |Sub_{k_{++}}(T^+_x) \setminus{Sub_{k_{++}}(T^+_y)\cup Sub_{k_{--}}(T^-_y)}|}$.%,
  }
  \label{fig:CDC_divide}
\end{figure}

\begin{lemma}\label{lem:ordersDescribing2D} 
For any distribution of a finite set of two dimensional points $V$ in two dimensional space, there are two total orders $T_x=(V,\leq_x)$ and $T_y=(V,\leq_y)$ such that for all $u,v\in V$ if $u_x \odot v_x$ then $u \odot_x v$ and if $u_y \odot v_y$ then $u \odot_y v$ and such that $|Sub_k(T_x)\cap Sub_k(T_y)| \geq k-1/3n$ for every $0 \leq k \leq n$.
\end{lemma}
\begin{proof}
Constructing orderings such that such that for all $u,v\in V$ if $u_x \odot v_x$ then $u \odot_x v$ and if $u_y \odot v_y$ then $u \odot_y v$ is trivial.
From here on, we assume that $n$ is a multiple of $3$.
Take the four total orders that can be generated by sorting our distribution in (1) either positive or negative and (2) either $x$- or $y$-direction.
Call these four total orders $T^+_x=(V,\leq^+_x)$, $T^-_x=(V,\leq^-_x)$, $T^+_y=(V,\leq^+_y)$ and $T^-_y=(V,\leq^-_y)$.
Let $k_{++}>n/3$ be the smallest integer such that $|Sub_{k_{++}}(T^+_x) \setminus{Sub_{k_{++}}(T^+_y)}| = n/3$ but $|Sub_{k_{++}+1}(T^+_x)\setminus{Sub_{k_{++}+1}(T^+_y)}| > n/3$, or if no such integer exists, then $k_{++}=n$.
Similarly we define $k_{--}$ for $T^-_x$ and $T^-_y$, $k_{+-}$ for $T^-_x$ and $T^+_y$ and $k_{-+}$ for $T^+_x$ and $T^-_y$.
By symmetry and renaming we can safely assume that $k_{++}$ is not smaller than any of $k_{+-}$, $k_{-+}$ and $k_{--}$. 
Using $Sub_{k_{++}}(T^+_x)$, $Sub_{k_{--}}(T^-_x)$, $Sub_{k_{++}}(T^+_y)$ and $Sub_{k_{--}}(T^-_y)$ we now partition $V$ into nine subsets and label these as shown in Figure~\ref{fig:CDC_divide}.
We will now prove that $k_{++} = n$ using proof by contradiction.
First we assume that $k_{++} < n$ but that $k_{++} + k_{--} \geq n$:

In this case we know that ${n-k_{++}\leq k_{--}}$, hence \[{|Sub_{n-k_{++}}(T^-_x) \setminus{Sub_{n-k_{++}}(T^-_y)}| \leq n/3}\] and that \[{|Sub_{n-k_{++}-1}(T^-_x) \setminus{Sub_{n-k_{++}-1}(T^-_y)}| \leq n/3}.\]
But \[Sub_{k_{++}}(T^+_x) \setminus{Sub_{k_{++}}(T^+_y)} = Sub_{n-k_{++}}(T^-_y) \setminus{Sub_{n-k_{++}}(T^-_x)}\] and \[{Sub_{k_{++}+1}(T^+_x) \setminus{Sub_{k_{++}+1}(T^+_y)}} = {Sub_{n-k_{++}-1}(T^-_y) \setminus{Sub_{n-k_{++}-1}(T^-_x)}}.\]
This leads to a contradiction since \[{n/3 \geq Sub_{n-k_{++}-1}(T^-_y) \setminus{Sub_{n-k_{++}-1}(T^-_x)} > n/3}.\]
Hence, our assumption must be incorrect.

In our second case we assume that $k_{++} < n$ but that $k_{++} + k_{--} < n$:
Our first observation is that ${x_{1,2}+x_{1,3} = x_{2,3}+x_{1,3} = n/3}$ by our definition of $k_{++}$.
For simplicity we will assume $x_{1,2} = x_{2,3}$ and similarly $x_{2,1} = x_{3,2}$.
Again, by arguments about symmetry we can safely assume that $x_{1,3} \geq x_{3,1}$.
We have that $x_{1,3}\geq n/6$, otherwise $x_{1,2}+x_{1,3}+x_{2,3}+x_{2,1}+x_{3,1}+x_{3,2} > n$.
By our definition of $K_{++}$, $x_{1,1}<n/3$ must hold, otherwise $Sub_{k_{++}}(T^+_x) \cup Sub_{k_{++}}(T^+_y) = V$ and $k_{++}=n$.
Similarly $x_{1,1}\geq 1$, otherwise ${x_{1,2}=x_{2,1}=x_{3,2}=x_{2,3}=n/3}$, which can only be true if $n=0$ and then $k_{++}=n$.
Assume that $x_{3,1} = x_{1,3} = n/3$.
This implies that $x_{1,1}+x_{2,2}+x_{3,3}=n/3$ and $x_{1,2}=x_{2,1}=0$ and a situation where $k_{+-}>k_{++}$, as for all $n/3\leq k \leq n$ then ${k - |Sub_{k}(T^+_x) \cap Sub_{k}(T^-_y)| \leq n/3}$.
Hence this also contradicts our original assumption of ${k_{++}\geq k_{+-}}$.
So $x_{3,1}<n/3$ and $x_{2,1}>1$ as $x_{2,1}+x_{3,1}=n/3$.
But this means that $x_{1,1}+x_{1,2}+x_{1,3}+x_{2,3}+x_{3,3}<n/3$ (recall that all $x_{i,i}$ are disjoint) which in turn implies that $Sub_{k_{++}}(T^-_x) \setminus Sub_{k_{++}}(T^+_y) < n/3$ and hence $k_{++}<k_{+-}$.
But $k_{++}\geq k_{+-}$ so we again have a contradiction!
Hence $k_{++}=n$ is the only viable alternative, completing our proof.    
\end{proof}

By Lemma~\ref{lem:ordersDescribing2D} the following result becomes fairly straightforward.

\begin{lemma}   \label{lem:cdc2}
{\cdc} can be solved in $\Ordo^*((cn)^{2n/3})$ time with dynamic programming.
\end{lemma}

\begin{proof}
From Lemma~\ref{lem:ordersDescribing2D} we know that there is a combination of directions such that we at most need to keep track of the exact ordering of $n/3$ points in each direction, yielding a factor $OBN(n/3)^2$.
However, Lemma~\ref{lem:ordersDescribing2D} hides a small detail in that the last set of variables added to our total ordering(s) may bring the total ordering above $n/3$ variables.
But as this only occurs for the very last part of the ordering, it only adds a factor $2^{O(n)}$.
On top of that the variables also needs to be partitioned into which are part of which of the two orderings, which have already been used, and which have not yet been used, giving a factor of $2^{O(n)}$.
Combined we hence have a complexity of $O((cn)^{2n/3})$ for some constant $c$.
\end{proof}

To improve this approach further we implement the same methods as we used in Section~\ref{sec:allen}:
instead of strictly using orderings we use records containing pairs $(S_i,B_i)$ where $S_i$ are subsets of $V$ and $B_i$ is the Boolean keeping track of if we can add new variables to the given set or not.
Similarly to the {\allen} case we also limit the number of such pairs we allow at once.
What differs from {\allen} to {\cdc} is that we, again, for {\cdc} need to keep track of two orderings, giving us both $(S^x_0,B^x_0),\dots,(S^x_m,B^x_m)$ and $(S^y_0,B^y_0),\dots,(S^y_m,B^y_m)$.
Hence the final form of the records are $(S, (S^x_0,B^x_0),\dots,(S^x_m,B^x_m),(S^y_0,B^y_0),\dots,(S^y_m,B^y_m))$.
Further, the notation of left-heaviness needs to be adapted to the use of two ordered partitions:
For all $i\in \{0,\dots,m\}$ every variable in some $S^{x}_i$ also needs to be in either $S^{y}_0$ or $S^{2}_1$ and every variable in $S^{y}_i$ also needs to be in either $S^{x}_0$ or $S^{x}_1$.
By also balancing $S^x_0 \cup S^y_1$ and $S^y_0 \cup S^y_1$ to remain of roughly similar size, we also keep the advantages from our earlier improvements to {\cdc}.  
Pair-placing here means that once we place a variable in $S^x_{1}$ or $S^y_{1}$ we also simultaneously place it in $S^y_{i}$ or $S^x_{i}$ respectively.
With this approach, combined with similar logic to our second improvement for {\cdc}, we at most need as many holes as the current number of unplaced variables.
If more is needed, the solution is easier to find in another orientation.

\begin{theorem}   \label{lem:cdc3}
{\cdc} can be solved in $\Ordo^*((\frac{cn}{\log{n}})^{2n/3})$ time using dynamic programming with sublinear partitioning.
\end{theorem}
\begin{proof}
(Sketch)
By using records of the form \[(S, (S^x_0,B^x_0),\dots,(S^x_i,B^x_i),(S^y_0,B^y_0),\dots,(S^y_j,B^y_j)),\] balancing $S^x_0 \cup S^y_1$ and $S^y_0 \cup S^y_1$ to roughly equal size, and using our substitute for left-heaviness and pair-placing, we can apply Algorithm~\ref{alg:allenMain} with only a few minor modifications.
While we in Lemma~\ref{lem:cdc2} effectively found the optimum for $|Sub_{k_{++}}(T^+_x) \setminus{Sub_{k_{++}}(T^+_y)}|!$ we are here instead interested in optimizing \[(2\cdot|V\setminus \{Sub_{k_{++}}(T^+_x) \cup {Sub_{k_{++}}(T^+_y)}\}|+1)^{|Sub_{k_{++}}(T^+_x) \setminus{Sub_{k_{++}}(T^+_y)}|}.\]
I.e., assuming there are $m$ unplaced variables we need at most $m$ holes in both the $x$ and $y$ direction, while partitioning our current variables into the roughly $2m+1$ possible sets in each direction, yielding $(2m+1)^{(1-1/3)n-m}$.
Since this function, for large $n$, has a smaller optimum than $n!$, our complexity is directly given by Lemma~\ref{lem:cdc2} combined with the same logic as for Theorem~\ref{theo:mainAllen}.
Similarly, correctness also follows via similar arguments to those  in Theorem~\ref{theo:mainAllen}.
\end{proof}

With Theorem~\ref{lem:cdc3} in place we now have an $\ordo(n)^{2n/3}$ algorithm for {\cdc}.

\section{Discussion and Conclusion}\label{sec:conc}
While the results presented in the paper yield significant improvement to the worst-case complexity of {\allen} and {\cdc}, a plethora of questions and open ends remain.

For {\allen} we consider two major open questions left by our results;
the first one being the true value of $c$ 
and the second that our approach with using left-heaviness seems overly restrictive.
Is there a better algorithm that does not necessitate the restriction of left-heaviness and instead allows merging sets to both the left and the right?
While removing the restriction of left-heaviness is not difficult, it significantly complicates the analysis, and we did therefore not attempt it.

The questions raised for {\allen} naturally also hold for {\cdc}.
However, in this case the value of the exponent is arguably of more interest.
While we saw in Section~\ref{sec:cdc} how the exponent $2n/3$ occurred naturally, many general problems solvable in $2^{\Ordo(n)}$ time often have $2^{\Ordo(\sqrt{n})}$ algorithms for the two-dimensional variants.
Could this imply that there might be an $\Ordo^*((cn)^{n/2})$ algorithm for {\cdc}?
Finding such an algorithm, and maybe even improving it further to $\Ordo^*((\frac{cn}{\log{n}})^{n/2})$ would be very interesting.

The main idea behind the dynamic programming with sublinear partitioning scheme is that we can dynamically restrict the amount of information stored. To the best of our knowledge, this approach has not been explored before in the literature, and we believe that it can be expanded further. 
For example, while we here focus on only one type of structure, there is nothing preventing us from storing multiple, e.g. exponentially many, different types of structures simultaneously.
This concept would likely be enough to at least match the $2^{\Ordo(n\log{\log{n}})}$ results for {\allen} restricted to only interval length one of~\cite{DBLP:conf/kr/DabrowskiJOO20}, while keeping the $\Ordo^*((\frac{cn}{\ln{n}})^n)$ complexity for general {\allen}.

In conclusion, we have in this paper seen how we can use structural properties required by solutions for {\allen} and {\cdc} instances to achieve, significantly, improved upper bounds for these problems.
In particular we have seen how we using dynamic programming combined with optimizing how much information we store in each iteration, can solve {\allen} in $\Ordo^*((\frac{cn}{\ln{n}})^n)$ time {\cdc} in $\Ordo^*((\frac{cn}{\ln{n}})^{2n/3})$ time.

\section*{Acknowledgements}
The first author is partially supported by the {\em National Graduate School in Computer Science} (CUGS), Sweden. The second author is partially supported by the Swedish Research Council (VR) under grant 2019-03690.

\bibliographystyle{abbrv}
\bibliography{bib}

\end{document}